\documentstyle[12pt]{article}
\def\RR{{\rm I\!R}} %real numbers
\def\one{{\mathchoice {\rm 1\mskip-4mu l} {\rm 1\mskip-4mu l}
	{\rm 1\mskip-4.5mu l} {\rm 1\mskip-5mu l}}}
\def\CC{{\mathchoice {\setbox0=\hbox{$\displaystyle\rm C$}\hbox{\hbox
	to0pt{\kern0.4\wd0\vrule height0.9\ht0\hss}\box0}}
	{\setbox0=\hbox{$\textstyle\rm C$}\hbox{\hbox
	to0pt{\kern0.4\wd0\vrule height0.9\ht0\hss}\box0}}
	{\setbox0=\hbox{$\scriptstyle\rm C$}\hbox{\hbox
	to0pt{\kern0.4\wd0\vrule height0.9\ht0\hss}\box0}}
	{\setbox0=\hbox{$\scriptscriptstyle\rm C$}\hbox{\hbox
	to0pt{\kern0.4\wd0\vrule height0.9\ht0\hss}\box0}}}}
\def\ZZ{{\mathchoice {\hbox{$\sans\textstyle Z\kern-0.4em Z$}}
	{\hbox{$\sans\textstyle Z\kern-0.4em Z$}}
	{\hbox{$\sans\scriptstyle Z\kern-0.3em Z$}}
	{\hbox{$\sans\scriptscriptstyle Z\kern-0.2em Z$}}}}
\font \fivesans  = cmss10 at 5pt
\font \sevensans = cmss10 at 7pt
\font \tensans   = cmss10
\newfam\sansfam
\textfont\sansfam=\tensans
\scriptfont\sansfam=\sevensans
\scriptscriptfont\sansfam=\fivesans
\def\sans{\fam\sansfam\tensans}
\newcommand{\hf}{{\textstyle\frac{1}{2}}}
\newcommand{\bw}{{\textstyle\bigwedge}}
\newcommand{\Wp}{{\textstyle\bigwedge^+}}
\newcommand{\Wm}{{\textstyle\bigwedge^-}}
\newcommand{\Om}{\Omega}
\newcommand{\Tr}{{\rm Tr\,}}
\newcommand{\tr}{{\rm tr\,}}
\newcommand{\DD}{I\kern-3.5pt D}
\newcommand{\FF}{I\kern-3.5pt F}
\newcommand{\rf}[2]{{\textstyle\sqrt{\frac{#1}{#2}}}}
\newcommand{\af}{{\textstyle\frac{1}{\sqrt{3}}}a^4}
\newcommand{\sA}{{\textstyle\frac{1}{\sqrt{3}}}A^4}

\newcommand{\qu}{{\textstyle\frac{4}{3}}}
\begin{document}
\thispagestyle{empty}
\vspace*{-2cm}
\rightline{hep-th/9801040}
\vspace{4cm}
\begin{center}
    {\LARGE\bf Superconnections and the Higgs Field}\\[2cm]
    {\large G.\ Roepstorff}\\[5mm]
    Institute for Theoretical Physics\\
    RWTH Aachen\\
    D-52062 Aachen, Germany\\
    email:\ roep@physik.rwth-aachen.de\\[2cm]
    {\bf Abstract}
\end{center}
\begin{quote}
Within the mathematical framework of Quillen, one interprets the Higgs field as part of the 
superconnection $\DD$ on a superbundle. We propose to take as superbundle the exterior 
algebra $\bw V$ obtained from a Hermitian vector bundle $ V$ of rank $n$ with structure group 
$U(n)$ and study the curvature $\FF=\DD^2$. The Euclidean action, at most quadratic in 
$\FF$ and invariant under gauge transformations, depends on  $n+1$ central charges. 
Spontaneous symmetry breaking is related to a nonvanishing constant scalar curvature in the ground
state, $\FF=L_c^2$, where $L_c$ is the Higgs condensate. The $U(1)$ Higgs model is 
nothing but the familiar Ginzburg-Landau theory, whereas the $U(2)$ Higgs model relates to
the electro-weak theory (with two Higgs doublets). The present formulation leads to the relation
$g^2=3g'^2$ for the coupling constants,
the formula $\sin^2\theta=1/4$ for the Weinberg mixing angle, and the ratio 
             $  m_{\rm W}^2:m_{\rm Z}^2:m_{\rm H}^2=3:4:12 $
for the masses of W$^\pm$, Z$^0$, and the Higgs boson. Experimentally observed
deviations are attributed to radiative corrections.

\end{quote}
\newpage
\section{Introduction}
It has been generally accepted that spontaneously broken local gauge symmetries provide 
the correct framework for understanding the electro-weak interactions of elementary
particles [1]. The mechanism that gives masses to the $W^\pm$, $Z^0$, and leptons
however needs the introduction of a doublet of scalar fields, the so-called Higgs field,
with many puzzling features, physically as well as mathematically. The concepts of the
Higgs field and the related Higgs mechanism, over the years, have triggered many 
investigations, either from the supersymmetry or the differential geometry point of view.
 
Most attempts were a response to the fact that the Lagrangian of the standard model contains 
a large number of free
parameters, among them various gauge coupling constants, the parameters of the Higgs
potential, coupling constants of matter fields, and the elements of the quark mixing matrix.
Some of these constants are expected to come out of some kind of symmetry breaking 
mechanism occurring in some yet unknown theory while others can be chosen at our will.
One therefore feels that at present one is actually dealing with an effective 
(low energy, long range) field theory
where only some degrees of freedom appear explicitly. Consequently, no explanation for most 
of the constants, chosen to fit the experimental data, is offered.

As a normal mathematical setting one would perhaps regard the theory of fibre bundles
[2][3] that emerged as a primary tool for studying Yang-Mills systems.
Then the question may be raised: is the Higgs field an object of geometry?
Below we shall briefly survey some of the attempts to extend the formalism of gauge theory to
Yang-Mills-Higgs systems before trying to give new answers. 

A popular approach to the problem of assigning a geometrical role to the Higgs field comes
under the heading {\em dimensional reduction\/}. Witten [4], Manton [5], and Fairlie [6] 
were first to provide interesting model theories in higher dimension.
The reduction technique has been taken up again and used as a guiding principle by other 
authors [7]. In its simplest version it uses one extra dimension, flat metric, and translational 
invariance. Thus, one starts from a Yang-Mills connection on the trivial principal bundle 
$\RR^{n+1}\times G$ with compact semisimple Lie group $G$, considers the splitting 
$dx^0 A_0+dx^i A_i$ ($i=1,\ldots,n$) of the connection 1-form,
and identifies the Higgs field with $A_0$. The drawback is twofold: (1) the gauge
field is not allowed to depend on the extra variable $x^0$, (2) the Higgs field is always in the 
adjoint representation.

As a precursor of Quillen's superconnection theory, one may regard Ne'eman's proposal [8]
to make use of the supergroup $SU(2|1)$ for an algebraically irreducible electro-weak
unification. Supergroups are formal objects obtained from super-Lie algebras
where commutators are replaced by supercommutators. At first, the model appeared to
suffer from spin-statistics complications. The final treatment with Sternberg [9] however
took full advantage of Quillen's formalism. Super-Lie algebras are also at the heart of
an attempt [10] to construct a renormalizable model of gravity as a broken gauge theory.

Another approach borrows from the framework of non-commutative geometry (NCG) [11] 
and leads to what has been called {\em algebraic Yang-Mills-Higgs theories\/} [12] with 
obvious links to the supergroup formalism. The idea is to
replace the exterior algebra of differential forms (the de Rham complex) by some
noncommutative $\ZZ_2$-graded differential algebra. To start with, one replaces
$C^\infty(M)$ by $A\otimes C^\infty(M)$, where $A$ is some matrix algebra together 
with a grading automorphism and $M$ is spacetime.
As there is a generalized notion of what should be called a connection, by a proper choice 
of the algebra $A$ one can accommodate a Higgs field in the connection, be it a multiplet or 
several multiplets.  By now, many versions of the NCG approach have appeared which
successfully reformulate the standard model.
In the Connes-Lott approach [13], $A=\CC\oplus\CC$ whereas the Mainz-Marseille
group (see [12] for details) prefers $A=M_2(\CC)$ with grading automorphism diag(1,-1)
in both cases. Recently, Okumura [14] proposed yet another formulation.
When calculating the curvature, these authors get different results which influence the
Weinberg angle, the Higgs mass and the quartic Higgs coupling. Therefore, the predictive
power of the NCG approach has come under intense scrutiny.

Last but not least there are attempts to add a fifth `discrete dimension' [15] to spacetime with 
possible relation to parity and chiral symmetry breaking. We feel, though we cannot prove, that such an
approach, once fully worked out, will provide but another reformulation of a specific model
within the territory of non-commutative geometry.

In 1985 D.Quillen described his concept of a superconnection [16] (see also [17]), thereby abandoning
the traditional $\ZZ$-grading (of the exterior algebra of differential forms) in favor of 
a $\ZZ_2$-grading, giving thus more freedom to constructions in (commutative) differential geometry.
Bundles carrying a $\ZZ_2$-graded structure are termed superbundles. Quillen aimed at the 
construction of invariants of a superbundle (Chern-Weil forms) and the definition of the Chern
character of a superconnection. 
A serious attempt to extend the formalism of gauge theories using Quillen's concept of
superconnections has been launched in 1990 by R.Coquereaux et.al.[18]. It still borrows
from the NCG formalism. So does the work of C.-Y.\ Lee [19] and  H.\ Figueroa et.\ al.\ [20].

In the present paper, we do not rely on the NCG approach but strictly follow the guidelines of Quillen and try to paint
a coherent picture of  $U(n)$ Higgs systems whose ground states have constant 
curvature. The role we assign to the Higgs field is similar to the one of the NCG approach.
But the choice of the superbundle in new to the best of our knowledge.
The formalism has applications to Ginzburg-Landau theory [21] and topological field 
theory [22]. We shall leave out that aspect here.
%----------------------------------------------------------------------
\section{Superbundles}

We assume that $M$ is an oriented Riemannian manifold
of dimension $m$ which may be arbitrary. Later we shall be more specific and think of
$M$ as the four-dimensional Euclidean spacetime with flat Levi-Civita connection.
We let $\Gamma$ denote the algebra of smooth complex functions on $M$ and, if
$B$ is some bundle with base $M$, write $\Gamma(M,B)$ for the space of smooth
sections $s:M\to B$.

Pure Yang-Mills theory starts from a principal $G$ bundle $P$ over $M$, 
and an invariant action functional on the set ${\cal A}$ of connections on $P$. 
The compact semisimple Lie group $G$ is called the {\em gauge group\/} 
of the theory. 
The set ${\cal A}$ is modelled on the vector space of gauge fields, i.e., 
the space of 1-forms $A$ taking values in the bundle
$$
                          {\rm ad\,}P= P\times_G {\bf g}
$$
where ${\bf g}$ is the Lie algebra of $G$, and $G$ acts on ${\bf g}$ via the adjoint
representation. A gauge field $A$ can locally be written as $dx^\mu A_\mu(x)$ with 
$A_\mu(x)\in{\bf g}$. The electro-weak theory, however, is based on the non-semisimple
gauge group $SU(2)\times U(1)$ und thus admits two independent gauge couplings.
In what follows the focus will be on the non-semisimple case. Though we work with the gauge 
group $U(n)$, we argue in favor of only {\em one\/} coupling constant.

The notion of a gauge transformation is more subtle. Though, locally,  gauge transformations
may be thought of as maps from the base manifold $M$ into the group $G$, they cannot
be extended globally to sections of $P$ (unless $P$ is a trivial bundle). Instead, gauge 
transformations are {\em bundle automorphisms\/} of $P$. Automorphisms commute with 
the group action on $P$ by definition. Let ${\cal G}={\rm Aut}(P)$ denote the group of 
bundle automorphisms of $P$. A more explicit description of ${\cal G}$, which is closer 
to the physicists' notion, uses sections of the adjoint bundle:
$$
          {\cal G}=\Gamma(M,{\rm Ad\,}P)\ ,\qquad {\rm Ad\,}P=P\times_GG\ .
$$
The bundle ${\rm Ad}(P)$ is the associated bundle whose fibres are copies of the group $G$.
But the group action on $G$ is the adjoint action. The Lie algebra of ${\cal G}$ can
now be easily constructed: ${\rm Lie}{\cal G}  =\Gamma(M,{\rm ad\,}P)$.      

From now on we shall assume that $P$ is a principal bundle with structure group $G=U(n)$. 
Let the complex vector space $\CC^n$ be equipped with the standard scalar product so 
that its group of automorphisms is $G$. We may construct the associated bundle 
$$
                   V=P\times_G \CC^n
$$
which is a complex Hermitian vector bundle of rank $n$ with structure group $G$.
It is always understood that $G$ acts on the right of $P$ and on the left of $\CC^n$, and
the notation $\times_G$ means that we identify $(pg,z)\sim (p,gz)$ for $p\in P$, 
$z\in \CC^n$, and $g\in G$. 

Since algebraic constructions on vector spaces carry over to associated bundles, we may 
consider the exterior algebra $\bw V$ which is a Hermitian vector bundle of rank $2^n$ 
acted upon by gauge transformations $u\in{\cal G}$ via the representation $\bw$, 
namely, at $x\in M$ we have
\begin{equation}
             \bw u(x) \ \colon\ \bw V_x \to \bw V_x, \qquad u(x)\in U(n)\ .
\label{rep}
\end{equation}
Recall now that a {\em superspace\/} is a $\ZZ_2$-graded vector space whose elements
are said to have even or odd degree (or parity). Likewise, a {\em superbundle\/} is a vector 
bundle whose fibers are superspaces. Furthermore, a {\em superalgebra\/} has a superspace
as underlying vector space, and a product that respects the $\ZZ_2$-grading.
The exterior algebra $\bw V$ is both a superbundle and a superalgebra with grading
$$
          \bw V =\bw^+ V\oplus\bw^- V,\qquad  
           \bw^\pm V =\sum_{(-1)^p=\pm 1}\bw^p V\ .
$$
Though the subbundles $\bw^\pm V$ have the same rank $2^{n-1}$, 
there exists no natural isomorphism between them. It will soon become apparent
that only a spontaneous symmetry breaking connects $\bw^+ V$ with $\bw^- V$.

The remainder of this section is devoted to reviewing basic facts about the representation $\bw$
of $G=U(n)$ on $\bw\CC^n$. Since it is irreducible on each subspace $\bw^k\CC^n$ 
of $\bw \CC^n$, the commutant
$$
                    \textstyle (\bw G)' =\{ \sum_{k=0}^n c_kP_k\,|\, c_k\in\CC\},
$$
where $P_k$ projects onto $\bw^k\CC^n$, is an abelian algebra. 
Particular elements $C\in(\bw G)'$ will enter the Euclidean action. If $C=\sum c_kP_k$,
we shall refer to the numbers $c_k$ as {\em central charges\/}.

Another consequence is that the representation $\bw$ of $G$ respects the 
$\ZZ_2$-grading of $\bw \CC^n$  and decomposes as $\Wp\oplus\Wm$.  We thus write
$$
                 \bw u =\pmatrix{\bw^+u & 0\cr 0& \bw^-u\cr} ,\qquad u\in U(n)\ .
$$
Because the operator $\bw u$ does not change the parity of vectors, it is said to be {\em even}.

Similar properties may be established for the induced representation $a\mapsto \hat{a}$ 
of the Lie algebra ${\bf g}={\bf u}(n)$ given by
$$
     \hat{a} =\frac{d}{dt}\bw\exp(ta)|_{t=0}
    =\pmatrix{\hat{a}^+ &0\cr 0&\hat{a}^-\cr}\  ,\qquad a\in{\bf u}(n),\qquad
   \hat{a}^\pm\in{\rm End\,}\bw^\pm \CC^n\ .
$$
In fact, $\hat{a}$ is the unique extension of $a\in{\rm End}\, \CC^n$ to an even derivation
of the algebra $\bw  \CC^n$, i.e.,
$$
       \hat{a}(z\wedge z')=\hat{a}z\wedge z' + z\wedge\hat{a}z',\qquad z,z'\in\bw  \CC^n
$$
In particular,
\begin{eqnarray*}
    \hat{a}z &=& 0\phantom{u}\qquad z\in\bw^0 \CC^n\cong\CC\\
    \hat{a}z &=& az\qquad z\in\bw^1 \CC^n\cong \CC^n\ .
\end{eqnarray*}
An operator $L$ on $\bw  \CC^n$ is even (odd) if it preserves (changes) parity. 
This gives ${\rm End\,}\bw  \CC^n$ the structure of superalgebra:
$$     
 {\rm End\,}\bw  \CC^n={\rm End}^+\bw  \CC^n\oplus{\rm End}^-\bw  \CC^n\ .
$$
Note that $\hat{a}\in{\rm End}^+\bw  \CC^n$.

Up to normalization there exists a unique bilinear form $q(a,b)$, or equivalently a quadratic
form $q(a)=q(a,a)$, on the Lie algebra ${\bf su}(n)$, known as the Killing form, which is 
invariant and nondegenerate. By contrast, the Lie algebra ${\bf u}(n)={\bf su}(n)\oplus
{\bf u}(1)$, where $n\ge 2$, has a two-parameter family of such forms (we require that they 
be positive definite) parametrized by $g$ and $g'$:
\begin{equation}
    q(a) =-\frac{n}{g^2}\tr\Big(a-\frac{1}{n}\tr a\Big)^2-\frac{1}{g'^2}(\tr a)^2\ ,
   \qquad a\in {\bf u}(n)\ .   \label{gg}
\end{equation}
When restricted to the subalgebra ${\bf su}(n)$, any member of this family 
reduces to a multiple of the Killing form as it should:
$$
     \tr a=0\ \ \Rightarrow\ \  q(a) \sim \tr {\rm ad}(a)^2=2n\,\tr a^2 \ .
$$
In the context of the electro-weak theory, $g$ and $g'$ are known as the two independent
gauge coupling constants. Unless one is committed to a specific representation of
the Lie algebra, there will be no a-priori relation between $g$ and $g'$. On the other hand,
given a distinguished faithful representation $\rho$, the condition $q(a)\sim -\tr\rho(a)^2$ 
fixes the ratio $g/g'$ once and for all.
                 
In fact, according to the point of view taken in this paper, the $U(n)$ Higgs model starts from a
distinguished representation, namely $\rho(a)=\hat{a}$, and thus provides a canonical choice
for the value of the ratio $g/g'$. Writing `Tr' for the trace on ${\rm End\,}\bw \CC^n$ 
(we reserve `tr' for traces in other circumstances) and setting
\begin{equation}
                 q(a) =-\Tr\hat{a}^2\ ,\qquad a\in{\bf u}(n)
    \label{q}
\end{equation}
we also fix the value of $g$ which is solely a matter of convenience without intrinsic meaning.

This choice of $q(a)$ may appear as an `article of faith'  and is certainly questionable; the matter is not 
being debated here. Instead, we will demonstrate how $g$ and $g'$ are related. 
Expanding both sides of the well-known formula 
$$
                  \log\Tr \exp\hat{a}=\tr\log(\one+e^a)\ ,
$$
we get:
\begin{equation}
            \Tr\one = 2^n\ ,\qquad
            \Tr\hat{a}=2^{n-1}\tr a\ ,\qquad 
            \Tr\hat{a}^2=2^{n-2}\big((\tr a)^2+\tr a^2\big)\ .   \label{Trace}
\end{equation}
Comparison with (\ref{gg}) shows that $(g/g')^2=n+1$. For the electro-weak theory
($n=2$), we get the equation $g^2=3g'^2$ apart from the value $g^2=2$ chosen
at will.

To prepare for later work, we introduce a basis $e_i$ ($i=1,\ldots,n^2$) in ${\bf u}(n)$ 
such that $q(e_i,e_k)=\delta_{ik}$. It is also assumed that $e_i$ ($i=1,\ldots,n^2-1$) 
is a basis for ${\bf su}(n)$.

Let us now investigate the two simplest situations: $U(1)$ and $U(2)$.
In the $U(1)$ case, it is obvious that 
$$
      \bw u =\pmatrix{1&0\cr 0&u\cr},\qquad \hat{a}=\pmatrix{0&0\cr 0&a\cr},\qquad
       u=e^a\in U(1),\ a\in i\RR\ .
$$ 
As expected, $e_1=i$ , since $q(i)=-i^2=1$.

We treat the $U(2)$ case in greater detail. Here $2^n=4$ and so $\bw u$
is a unitary $4\times 4$-matrix which is block diagonal (blocks are $2\times 2$ matrices):
$$
    \bw u =\pmatrix{\bw^+u & 0\cr 0 &\bw^-u\cr},\qquad
    \bw^+u=\pmatrix{1 &0\cr 0 & {\rm det\,}u\cr}, \qquad\bw^-u=u\in U(2)\ .
$$
It follows that
$$
      \hat{a} =\pmatrix{\hat{a}^+ &0\cr 0&\hat{a}^-\cr},\qquad
     \hat{a}^+=\pmatrix{0 &0\cr 0 & {\rm tr\,}a\cr}, 
      \qquad\hat{a}^-=a\in{\bf u}(2)
$$      
and $q(a)=-(\tr a)^2-\tr(a^2)$ by (\ref{q}) and (\ref{Trace}). Setting
$$
      a =a^k e_k =\frac{i}{\sqrt{2}}\pmatrix{\af+a^3      & a^1-ia^2\cr
                                                                                        a^1+ia^2 &  \af-a^3\cr}  
\qquad (a^k\in\RR) 
$$
we obtain a basis in {\bf u}(2) with the required property $q(a)=\sum_k(a^k)^2$.
%----------------------------------------------------------------------------
\section{Differential Forms}

We let $T^*M$ denote the complexified cotangent bundle of the manifold $M$.
Elements of $T^*M$ are said to be (complex) 1-forms.
The exterior algebra $\bw T^*M$ is a superbundle and so is $\bw(T^*M\oplus  V)$,
the algebra of $V$-valued differential forms.
The latter construction relates to the dimensional reduction formalism in an obvious way.
For, if $N$ is another manifold and $V=T^*N$,  then 
$T^*M\oplus  V\cong  T^*(M\times N)$.
Whether or not some manifold $N$ of dimension $n$ is lurking behind the scene,
there exists a natural isomorphism
\begin{equation}
                      \bw(T^*M\oplus  V) \cong \bw T^*M\otimes\bw V
   \label{iso}
\end{equation}
between $\ZZ_2$-graded algebras. The tensor product on the right hand side of (\ref{iso})
is special for graded algebras. It is often called a {\em skew tensor product\/}.
Generally speaking, if $X$ and $Y$ are $\ZZ_2$-graded algebras, the multiplication in $X\otimes Y$ is
given by
$$
               (x\otimes y)(x'\otimes y)=(-1)^r xx'\otimes yy',\qquad
                r=\cases{1 & if $x'$ and $y$ are odd\cr 0 & otherwise.\cr}
$$
The fact that the isomorphism (\ref{iso}) respects the grading means:
$$
      \bw^\pm(T^*M\oplus  V) \cong 
     \sum_{(-1)^{p+q}=\pm 1}\bw^p T^*M\otimes \bw^q V       
$$
or stated equivalently:
\begin{eqnarray*}
                 \Wp (T^*M\oplus  V) &\cong&
                 (\Wp T^*M\otimes\Wp V )\oplus(\Wm T^*M\otimes \Wm V ) \\
                \Wm(T^*M\oplus  V) &\cong&
                 (\Wm T^*M\otimes\Wp V) \oplus (\Wp T^*M\otimes \Wm V ) 
\end{eqnarray*}
The grading carries over to the space of $\bw V$-valued differential forms:
$$
      \Om :=\Gamma(M,\bw T^*M\otimes\bw V)=\Om^+\oplus\Om^-
$$
and to the algebra of sections of the endomorphism bundle:
\begin{equation}
     {\cal B} :=\Gamma(M,\bw T^*M\otimes{\rm End\,}\bw V)
                     ={\cal B}^+\oplus {\cal B}^-                                         \label{cala}
\end{equation}
Note that ${\rm End\,}\bw V$ is a superbundle, and the tensor product in (\ref{cala}) is
between graded algebras.
Elements of $A\in{\cal B}$ act on $\Omega$ and are called {\em local operators\/}
since they leave fibres intact. Equivalently, they commute with the multiplication by functions 
$f\in \Gamma$.

A local operator $A\in{\cal B}^\pm$ is said to have parity $\pm 1$ --- or is 
referred to as an even (odd) operator --- where parity is defined as follows:
\begin{eqnarray*}
 {\rm par}(A)=+1  \quad &\Leftrightarrow&\quad A\Omega^\pm\subset\Omega^\pm\\
 {\rm par}(A)= -1  \quad &\Leftrightarrow&\quad A\Omega^\pm\subset\Omega^\mp
\end{eqnarray*}
A different decomposition of $\cal B$ arises from the $\ZZ$-grading of $\bw T^*M$:
$$
    {\cal B}=\sum_{p=0}^m{\cal B}^p\ ,\qquad
    {\cal B}^p=\Gamma(M,\bw^pT^*M\otimes{\rm End\,}\bw V)
$$
Notice that $1\otimes {\rm id}$ serves as the unit in the algebra ${\cal B}$ and that there
are two natural embeddings:
\begin{eqnarray*}
   \Gamma(M,\bw T^*M) &\to& {\cal B}\ ,\ \omega\mapsto \omega\otimes{\rm id}\\
   \Gamma(M,{\rm End\,}\bw V ) &\to& {\cal B}\ ,\ A\mapsto 1\otimes A
\end{eqnarray*}   
Owing to these embeddings, various constructions on $\Gamma(M,\bw T^*M)$
and $\Gamma(M,{\rm End\,}\bw V )$ have extensions to $\cal B$.  

For instance, the operator of exterior differentiation $d$ on $\Gamma(M,\bw T^*M)$ 
may  be extended. No confusion will arise when we write $d$ in place of $d\otimes{\rm id}$.

The trace $\Tr :\Gamma(M,{\rm End\,}\bw V )\to \Gamma$ can be extended  
in an obvious manner:
$$
   \Tr\ :\ \Gamma(M,\bw T^*M\otimes{\rm End\,}\bw V)\ \to\ \Gamma(M,\bw T^*M)
   \ , \ \omega\otimes A \mapsto \omega\Tr A
$$
Any local operator $A$ may be decomposed into homogeneous components ($p$-forms):
$$
                A  = A_{[0]}+A_{[1]}+A_{[2]}+\ldots  ,\qquad   A_{[p]}\in {\cal B}^p
$$
The series truncates at $p=m$ where $m$ is the dimension of the manifold $M$, and
taking the trace of the top form, the integral 
\begin{equation}
                          {\rm Int}( A) = \int_M {\rm Tr\,}A_{[m]}\ \in\CC  \label{trace}
\end{equation} 
assigns complex numbers to local operators of compact support.

The Hodge star operator on $\Gamma(M,\bw T^*M)$ can uniquely be extended to a
real-linear operator on ${\cal B}$ so as to satisfy
\begin{equation}
                        *(AB)=B^**A,\qquad  A\in{\cal B},\ B\in{\cal B}^0\ .  \label{hod}
\end{equation}
Let $e_i$ ($i=1,\ldots,m$) be an oriented frame of the tangent bundle and $e^i$ the dual 
frame of $T^*M$. For any multi-index $I\subset\{1,\ldots,m\}$ we form the exterior product
\begin{equation}
     e^I =e^{i_1}e^{i_2}\cdots e^{i_p}, \qquad I=\{i_1, i_2,\ldots,i_p\},\quad
     i_1<i_2<\cdots<i_p,\quad p=|I|                   \label{frame}
\end{equation}
to obtain a frame of $\bw T^*M$. It is assumed that $e^{\emptyset}=1$.
Using (\ref{hod}), we have 
\begin{eqnarray*}
      *(e^I\otimes A_I) &=& *\big((e^I\otimes {\rm id})(1\otimes A_I)\big)\\
                                     &=&(1\otimes A_I^*)(*e^I\otimes{\rm id}) \\
                                     &=&(\pm 1)^{m-p}*e^I\otimes A_I^*\ , 
           \qquad A_I\in\Gamma(M,{\rm End}^\pm\bw V).
\end{eqnarray*}
Let $d\tau=*1=e^1e^2\cdots e^m$ denote the volume element. Then there are
functions $g^I\in\Gamma$ such that
$$
        e^I*e^J =\cases{g^I d\tau & if $I=J$\cr 0 & if $I\ne J$\cr}
$$
and with reference to (\ref{frame}): $ g^I=(e^I,e^I)={\rm det\,}(e^{i_k},e^{i_l})_{k,l
=1,\ldots,p}>0$. The algebra ${\cal B}$ may be equipped with a scalar product,
\begin{equation}
                            (A,B)={\rm Int}(B*A),\qquad A,B\in{\cal B}\ ,  \label{prod}
\end{equation}   
and, by a straightforward calculation,
$$
         \|A\|^2=(A,A)=\int_M d\tau\,\sum_I g^I \Tr(A_IA_I^*) \ge 0\ ,\qquad
           A=\sum_I e^I A_I
$$
The norm $\|\cdot\|$ on ${\cal B}$ will be used in Section 5 to construct the Euclidean
action.

%-----------------------------------------------------------------------------
\section{Superconnections}

We start with a few remarks about connections. With $P$ a principal $G$ bundle,
where $G=U(n)$, the space ${\cal A}$ of connections is an affine space with non-trivial
topology if $n\ge 2$, e.g. $\pi_0({\cal A})=\ZZ$. With
$\cal G$ acting on ${\cal A}$, it seems natural to pass to the quotient
$$
                           B{\cal G} ={\cal A}/{\cal G}
$$
to obtain the classifying space for $\cal G$ bundles. In physics, $B{\cal G}$ is known as the
space of gauge orbits. It corresponds to the phase space of classical mechanics.
The passage to statistical mechanics is mirrored, in Euclidean field theory, by the
process of quantization, i.e., the introduction of path integrals over ${\cal A}$. 
Granted the absence of anomalies, path integrals project onto $B{\cal G}$. The calculation, 
however, requires gauge fixing and the introduction of Faddeev-Popov ghosts. Gribov's 
discovery, also known as the {\em Gribov ambiguity\/}, may be rephrased by saying that there 
is no continuous global choice of gauge or, stated more formally, ${\cal A}$ does not
admit a smooth global section. Though these intricacies are not the subject of the present paper,
we should be aware that some of the formulas below hold only on local
coordinate patches without explicit mentioning.

The advantage of giving a connection $A$ on the principal bundle $P$ is that it determines a
connection on every associated bundle and thus provides covariant derivatives $d_A$
on various vector bundles. We use the terms {\em connection\/} and {\em covariant 
derivative\/} interchangeably. A connection on the bundle $V$ simply is a linear map
$$
                              d_A\ \colon\ \Gamma(M, V)\to\Gamma(M,T^*M\otimes V)
$$
satisfying the Leibniz rule $d_A(fs)=df\, s+ fd_As$ for all functions $f$ and sections $s$.
The connection extends in a unique way to an operator $d_A$ on $\Gamma(M,\bw T^*M
\otimes V)$ sending $p$-forms to $(p+1)$-forms. Locally, $d_A =d+A$ where 
$A\in \Gamma(M,T^*M\otimes{\rm End\,}V) $ is the connection 1-form or gauge field.
The 2-form $F=d_A^2\in\Gamma(M,\bw^2T^*M\otimes{\rm End\,}V)$ is said to be
the {\em curvature\/} of the connection $d_A$. In terms of physics, $F$ is the field 
strength of a gauge theory. Under a gauge transformation,
$$
   {\phantom |}^u\kern-1pt  A=uAu^{-1}+u\,du^{-1},\qquad
   {\phantom |}^u\kern-1pt  F=uFu^{-1}\ ,\qquad u\in{\cal G}\ .
$$
We may pass now to the superbundle $\bw V$ and lift the fields $A$ and $F$ to certain local 
operators on $\Omega$ of definite parity:
$$
       \hat{A}=\pmatrix{\hat{A}^+&  0\cr  0 &\hat{A}^-\cr}\in{\cal B}^-,\qquad
       \hat{F}=\pmatrix{\hat{F}^+ & 0\cr  0 &\hat{F}^-\cr}\in{\cal B}^+
$$
Of course, $\hat{A}$ and $\hat{F}$ are still  one- and two-forms respectively. Recall that
the matrix representation refers to the $\ZZ_2$-grading of $\bw V$.
In the same manner, $d_A$ can be lifted to a connection $D$ on the superbundle:
$$
           D =\pmatrix{ D^+ &0\cr 0 & D^-\cr}\ ,\qquad \hat{F}=D^2\ .
$$
Locally, we have $D=d+\hat{A}$ and $D^\pm =d+\hat{A}^\pm$. When acting on 
$\Omega$, the differential operator $D$ changes the parity and so is of odd type. 

To extend the connection $D$ to a superconnection $\DD=D+L$ we introduce a skew
selfadjoint operator $L$ on $\Gamma(M,\bw V)$ of odd type,
$$
                            L=\pmatrix{0 & i\Phi^*\cr i\Phi & 0\cr},
$$
formally a section of the bundle 
$$
                 \bw^0T^*M\otimes{\rm End}^-\bw V\cong {\rm End}^-\bw V\ .
$$
and hence an element of ${\cal B}^-\cap{\cal B}^0$.
The complex scalar field $\Phi(x)$ is said to be the {\em Higgs field\/} of the system. 
It has the following characteristic properties:
\begin{itemize}
\item At $x\in M$, the Higgs field $\Phi(x)$ is a linear map from 
          $\bw^+V_x$ to $\bw^-V_x$. Consequently, $\Phi^*(x)$ maps 
          $\bw^-V_x$ to $\bw^+V_x$.
\item Under a change of the gauge,
    $$
    {\phantom |}^u\Phi =(\Wm u)\Phi (\Wp u)^{-1},\quad 
    {\phantom |}^u\Phi^* =(\Wp u)\Phi^* (\Wm u)^{-1}         
    $$
    which is summarized by 
   $$
                            {\phantom |}^uL=(\bw u)L(\bw u)^{-1}\ .
   $$
\item Like any section of the bundle ${\rm End}^-\bw V$,  $L$ extends to an odd operator 
   on $\Omega$. In more detail: $L$ acts on $\bw V$-valued $p$-forms by
   $$
        L\big(dx^{\mu_1}\wedge\cdots\wedge dx^{\mu_p}\,v_{\mu_1\cdots\mu_p}(x)\big)
    =(-1)^p dx^{\mu_1}\wedge\cdots\wedge dx^{\mu_p}L(x) v_{\mu_1\cdots\mu_p}(x)
   $$
   so as to be in accord with the skew tensor product $\bw T^*M\otimes {\rm End}\bw V$. 
   To put it differently, $L$ satisfies the rule $\{L,dx^{\mu}\}=0$ or, equivalently, $L$ 
   anticommutes with the multiplication by $\Gamma$-valued 1-forms. Thus $L : 
   \Omega^\pm\to\Omega^\mp$ is parity changing, hence $L\in{\cal B}^-$ by construction.
\item Since both $d$ and $L$ are odd degree operators, their anticommutator 
  (or supercommutator) $dL:=\{d,L\}$ is an even operator (and a 1-form) called the 
  covariant derivative of $L$. Similarly, the anticommutator 
   $$
                       DL :=\{D,L\}=\pmatrix{0 & i(D\Phi)^*\cr iD\Phi & 0\cr}
   $$
   provides the covariant derivatives of the Higgs field and its adjoint:
   \begin{eqnarray*}
      D\Phi := D^-\Phi+\Phi D^+ &=& d\Phi+\hat{A}^-\Phi+\Phi \hat{A}^+  \\
              &=& dx^\mu(\partial_\mu\Phi+\hat{A}_\mu^-\Phi-\Phi \hat{A}_\mu^+)\\
     (D\Phi)^*:= D^+\Phi^*+\Phi^* D^- 
             &=& d\Phi^*+\hat{A}^+\Phi^*+\Phi^* \hat{A}^-  \\
             &=& dx^\mu(\partial_\mu\Phi^*+\hat{A}_\mu^
                       +\Phi^*-\Phi^* \hat{A}_\mu^-)
   \end{eqnarray*}
     Here, we used the fact that $\Phi$ and $\Phi^*$ anticommute with $dx^\mu$.
\end{itemize}
Finally, the operator 
$$
                \DD=D+L=\pmatrix{D^+ & i\Phi^*\cr i\Phi & D^-\cr}
$$
defines a {\em superconnection\/} on the superbundle $\bw V$ in the sense of Quillen: 
$\DD$ is a differential operator of odd type on $\Omega$, hence acts on $\bw V$-valued 
differential forms. It no longer sends $p$-forms to $(p+1)$-forms,
but sends odd elements of $\Omega$ to even elements and vice versa so as to 
satisfy a Leibniz formula.

In physics, fields are viewed as varying objects. Varying the gauge field means 
passage from one superconnection $\DD$ to another, say $\DD'$, such that
the difference $\DD-\DD'$ comes out as a local operator
built upon 1-forms (the diagonal parts) and 0-forms (the off-diagonal parts). 
Hence the notion of a {\em superconnection on a superbundle\/} 
is in accordance with the requirement that, whatever the context, connections form an 
affine space modelled on some set of local operators. 

From $\FF=(D+L)^2=D^2+\{D,L\}+L^2$ we obtain the decomposition
$$
                            \FF =\FF_{[0]}+\FF_{[1]}+\FF_{[2]}\in{\cal B}^+,\qquad
                            \FF_{[p]}\in{\cal B}^p
$$
for the curvature $\FF$ of the superbundle $\bw V$.
In particular, the curvature is a local operator (not a differential operator).
Note that the Bianchi identity $[\DD,\FF]=0$ is a trivial consequence of the 
definition of $\FF$.

As indicated, the curvature $\FF$ has homogeneous components for $p=0,1,2$.
The 0-form is bilinear in the Higgs field,
$$
          \FF_{[0]} =L^2=\pmatrix{-\Phi^*\Phi & 0\cr
                           0&-\Phi\Phi^*\cr}\ ,
$$ 
while the 1-form is linear in the covariant derivatives of the Higgs field:
$$
          \FF_{[1]} = DL = \pmatrix{0 & i(D\Phi)^*\cr iD\Phi &0\cr}
$$
Finally, the 2-form
$$
                   \FF_{[2]} =D^2=\hat{F}=\pmatrix{ \hat{F}^+ & 0\cr 0 & \hat{F}^-\cr}
$$
gives the curvature when the Higgs field is absent.

 %----------------------------------------------------------------------
\section{Euclidean Action and Stationary Points}

We shall always stay within the realm of Euclidean field theory. For the remainder
of this paper, $M$ denotes the four-dimensional Euclidean flat spacetime
with standard orientation: $d\tau=dx^1\wedge dx^2\wedge dx^3\wedge dx^4$.

Before describing the field equations of the Higgs model, we motivate the construction of
a gauge invariant Euclidean action based on the superbundle $\bw V$ and the gauge group 
$U(n)$. With $F$ the curvature of a Yang-Mills connection, one takes $S=\hf\|F\|^2$ 
as the action so that the global minimum is attained for the flat connection. Similarly,
the superbundle is flat if $\FF=0$. However, the definition $S=\hf\|\FF\|^2$ gives
us models that show no sign of spontaneous symmetry breaking. To our rescue comes the
abelian algebra $(\bw G)'$ of gauge invariant operators $C$, each of them constant on $M$.
If $C$ is selfadjoint, the following definition of the Euclidean action serves the purpose:
\begin{equation}
            S =\hf\|\FF+\mu^2C\|^2  \ ,\qquad C\in(\bw G)',
           \label{ea}
\end{equation}
Euclidean actions that differ by the choice of $C$ are said to be {\em phases\/} of the same 
model. As an element of an abelian algebra, $C$ can always be written in terms of central 
charges $c_k$, $k=0,\ldots,n$.  Selfadjointness of $C$ makes these charges real numbers.
We may write
$$
   C =\pmatrix{C^+ & 0\cr 0 & C^-},\qquad C^\pm =\sum_{(-1)^k=\pm 1}c_kP_k
$$
and split the action into different parts for easier interpretation,
\begin{equation}
          S=\hf\|\hat{F}\|^2+\hf\|DL\|^2+\hf\|L^2+\mu^2C\|^2\ .   \label{hl}
\end{equation}
The last term involves the Higgs potential $V(\Phi)$:
\begin{eqnarray}
   \hf\|L^2+\mu^2C\|^2 &=&\int_M d\tau\,V(\Phi) \nonumber\\ 
   V(\Phi) &=& \hf\Tr (L^2+\mu^2C)^2\nonumber \\
                &= &\hf\Tr(\Phi^*\Phi-\mu^2C^+)^2     
                       +\hf\Tr(\Phi\Phi^*-\mu^2C^-)^2   \label{HiPo}
\end{eqnarray} 
In (\ref{hl}) we encounter the term $\hf\|\hat{F}\|^2$ as part of the action.
To analyze it we introduce the components of the curvature $F$ with respect to
the basis $dx^\mu$  in $T^*M$ and the basis $e_k$ in ${\bf u}(n)$:
$$
            F =F^k e_k,\qquad  F^k=\hf dx^\mu\wedge dx^\nu F_{\mu\nu}^k(x)
$$
The coefficients $F_{\mu\nu}^k(x)$ are real functions on $M$.
It follows that $\hat{F} =-\hat{F}^*=F^k \hat{e}_k$ and, in view of the definitions 
(\ref{prod}) and (\ref{trace}),
$$
            \hf\|\hat{F}\|^2 
             =\int_M d\tau\,{\textstyle\frac{1}{4}}\sum_{k,\mu\nu}(F^k_{\mu\nu })^2
$$
This reveals that $\hf\|\hat{F}\|^2$ is the Euclidean action, correctly normalized,
of a conventional gauge theory without Higgs field.

The positive parameter $\mu$ sets the mass scale of the model while the central charges $c_k$
control the expectation value of the Higgs field (the so-called condensate) on the 
classical (or tree) level. The role of these parameters is similar when the theory is quantized 
using path integrals. The latter procedure replaces the classical Higgs potential by an
effective potential whose minima describe the vacuum states.

One remark is in order. Recall that we have deliberately put $g^2$ equal to 2.
It may later be necessary to work with an arbitrary value of the gauge coupling $g$.
The introduction of $g$ as a parameter can be achieved by an appropriate scaling:
$$
        S\ \to\ \lambda^{-2}S(\lambda A,\lambda\Phi),\qquad 2\lambda^2=g^2.
$$
Since scaling has no influence on second order terms of the action, it will not alter the results
of the present paper. 

Let us write the kinetic term of the action involving the Higgs field in more conventional terms:
$$
    \hf\|DL\|^2   = \int_M d\tau\,\sum_\mu\Big(\Tr(\partial_\mu\Phi^*\partial_\mu\Phi)
                               +A^k_\mu j_{k\mu}\Big)
$$
where $d\Phi=dx^\mu\partial_\mu\Phi(x)$ and $A=dx^\mu A_\mu^k(x)e_k$.
To each component $A^k$ we associated a current 
$$
                j_k=dx^\mu j_{k\mu}(x)=\Tr(\hat{e}_k\{L,DL\})\qquad(k=1,\ldots,n^2)\ .
$$      
Even more explicitly, we have 
$$
                \{L,DL\} =\pmatrix{J^+&0\cr 0&J^-}
$$
where
\begin{eqnarray*}
  J^+ &=& -\Phi^*D\Phi-(D\Phi)^*\Phi\ \ =\ 
              dx^\mu\big(\Phi^*(D\Phi)_\mu-(D\Phi)_\mu^*\Phi\big)\\
  J^- &=& -\Phi(D\Phi)^*-(D\Phi)\Phi^*\ =\ 
              dx^\mu\big(\Phi(D\Phi)^*_\mu-(D\Phi)_\mu\Phi^*\big) 
\end{eqnarray*}
Let us now consider the formal adjoint operators $d^*$, $d_A^*$, and $D^*$.
They belong to the standard repertoire of Yang-Mills systems.
Adjoints are always formed with respect to the scalar product of sections. 
Each of the above adjoint operators maps $p$-forms into 
$(p-1)$-forms. The operator $\delta:=-d^*$ is called the 
{\em coderivative\/} and $\Delta=-\{d,d^*\}$  the {\em Laplacian\/}. 

It proves convenient to write the Lie algebra valued current as $j=\sum_k j_k e_k$ so that
$\hat{j}=\sum_k j_k \hat{e}_k$.
From the condition that the action $S$ be stationary one obtains the field equations of the 
$U(n)$ Higgs model:
$$
        D^*\hat{F}+\hat{j}=0,\qquad  D^*DL=\{L,L^2+\mu^2C\}
$$
For this and similar calculations, it is useful to keep in mind that $D^*\hat{F}$ is short
for $ [D^*,\hat{F}]$ and $D^*DL$ is short for $[D^*,\{D,L\}]$. 

The field equations may also be put into a form reminiscent of previous Yang-Mills-Higgs 
models:
\begin{equation}
              d^*_A F+j=0 ,\qquad
              D^* D\Phi =  -2\Phi\Phi^*\Phi+\mu^2(\Phi C^+ + C^-\Phi )  
     \label{YMH}
\end{equation}
Again, $d_A^*F$ is short for $[d^*_A,F]$, and $D^*D\Phi$ is obtained from 
$$
     [D^*,\{D,L\}]=\pmatrix{ 0 & i(D^*D\Phi)^*\cr iD^*D\Phi &0\cr}\ .
$$
Any solution  $(A,\Phi)$ of the second order field equations (\ref{YMH}) is said to be a stationary point 
of the action. We should be aware that not every stationary point corresponds to a 
local or global minimum of the action.

The global minimum is attained if $A=0$ and if $L$ solves of the variational problem
\begin{equation}
                       \Tr(L^2+\mu^2C)^2 =\mbox{minimum.}         \label{vac}
\end{equation}
The solutions are said to describe (classical) vacua or ground states. Granted that $M$ is
connected, any solution $L_c$ of (\ref{vac}) is constant on $M$ and is referred to as 
the {\em Higgs condensate}. The group $U(n)$ acts upon the set of solutions, though not 
always freely: the residual gauge group 
\begin{equation}
                G_0 =\{u\in U(n)\ |\  (\bw u)L_c=L_c(\bw u)\}
\end{equation}
may well be nontrivial. Ground states that lie on the same gauge orbit are physically
equivalent. We must not expect the group $U(n)$ to act transitively: there may exist
many gauge orbits.

If $L_c^2$ is unique, we obtain a $U(n)$ invariant vacuum. Nonuniqueness is characteristic 
of a broken phase. Note also that each ground state has constant scalar curvature, 
$\FF=L_c^2$. The structure of $L_c$ is that of a constant matrix:
$$
   L_c=i\pmatrix{0&\Phi_c^*\cr \Phi_c& 0\cr} \ ,\quad \Phi_c=\mu v\ .
$$
There are two special cases where the variational problem (\ref{vac}) can be solved with
ease. First, $C=0$ implies $L_c=0$ giving a $U(n)$ invariant vacuum. Second, $C=1$
implies $L_c^2=\mu^2$, hence $v^*v=vv^*=1$, and $v$ establishes an isomorphism 
between the spaces $\bw^+\CC^n$ and $\bw^-\CC^n$. Conversely, any isomorphism 
$v$ gives us a solution of (\ref{vac}).

Suppose we look for exitations from some ground state, but ignore 
the Higgs degrees of freedom. Then $L$ is kept constant, i.e., $L=L_c$ and 
$DL=\{\hat{A},L_c\}=dx^\mu[\hat{A}_\mu,L_c]$. Provided $[\hat{A}_\mu,L_c]\ne 0$, 
the gauge particles acquire masses. Indeed, the mass term of the action may be written 
$$
     \hf\|DL_c\|^2=\int_M d\tau\,\hf\sum_\mu Q(A_\mu)\ ,
$$
where $Q$ is a positive semidefinite quadratic form on the Lie algebra:
\begin{equation}
     Q(a) =-{\Tr}[\hat{a},L_c]^2=a^i a^k m^2_{ik},\qquad a=a^i e_i\in{\bf u}(n)\ .
     \label{Qa}
\end{equation}
The eigenvalues of the matrix $(m^2_{ik})$ are the masses (squared) of gauge 
fields given by the eigenvectors where it is assumed that the eigenvectors are orthonormal
with respect to the bilinear form $q(a,b)$ on the Lie algebra obtained from (\ref{q}). 

Suppose now that $L_c'$ is another Higgs condensate giving rise to the quadratic 
form $Q'(a)$. Both $L_c$ and $L_c'$ lie
on the same gauge orbit if $(\bw u)L_c=L_c'(\bw u)$ for some $u\in U(n)$. Owing to the
invariance property $Q(a) =Q'(uau^{-1})$, the eigenvalues of the mass matrix
stay constant along any gauge orbit.

%------------------------------------------------------------------------------------
\section{The $U(1)$ Higgs Model}

A very simple situation arises when $n=1$ since there is only one basis element
$e_1=i$ in ${\bf u}(1)=i\RR$. We may thus write $A=idx^\mu A_\mu(x)$ and $F=i\hf
dx^\mu\wedge dx^\nu F_{\mu\nu}(x)$ with real-valued components $A_\mu$ and
$F_{\mu\nu}$. In the two-dimensional cap representation of ${\bf u}(1)$ we have 
$$
   \hat{A}=\pmatrix{0&0\cr 0&A\cr}\ ,\qquad \hat{F}=\pmatrix{0&0\cr 0&F\cr}
$$
The Higgs field is simply some complex scalar field $\Phi$. With $c_0$ and $c_1$ the
central charges, the Higgs potential becomes
\begin{eqnarray*}
    V(\Phi) &=&\hf(|\Phi|^2-\mu^2c_0)^2+\hf(|\Phi|^2-\mu^2c_1)^2\\[2mm]
                 &=& (|\Phi|^2-\mu^2c)^2
                          +{\textstyle\frac{1}{4}}\mu^2 (c_0-c_1)^2\ ,\qquad c=\hf(c_0+c_1)
\end{eqnarray*}
Provided that $c>0$, the minimum is attained for $\Phi=\mu ce^{i\alpha}$. 
Otherwise, the minimum is attained for $\Phi=0$.
There is no restriction in assuming that $c_0=c_1$ and $c=\pm 1$. From
$$
    \DD =\pmatrix{d & \Phi^*\cr -\Phi & d+A\cr}\ ,\qquad
    \FF  = \pmatrix{-|\Phi|^2 & i(d_A\Phi)^*\cr id_A\Phi & F-|\Phi|^2\cr}
$$
we obtain the action of the Ginzburg-Landau theory,
$$
    S =\int_M d\tau\,\Big({\textstyle\frac{1}{4}}\sum_{\mu\nu}F_{\mu\nu}^2
                                 +\sum_\mu |(\partial_\mu+iA_\mu)\Phi|^2
                                 +(|\Phi|^2-\mu^2c)^2\Big)\ ,
$$
whose current is given by
$$
      j=i dx^\mu j_\mu(x),\qquad  
      j_\mu=2\,{\rm Im}\big(\Phi^*(\partial_\mu+iA_\mu)\Phi\big)\ .
$$
For $c=1$, the system is in the superconducting phase, the residual gauge group is trivial,
and ground states differ by a constant 
phase: $\Phi(x)=\mu e^{i\alpha}$. However, these states belong to a single gauge orbit
and hence are equivalent. Provided $\Phi$ is kept at its ground state value, $S$ reduces 
to the action of a massive photon ($m_\gamma^2=2\mu^2$):
$$
          S =\int_M d\tau\,\Big({\textstyle\frac{1}{4}}\sum_{\mu\nu}F_{\mu\nu}^2
                                 +\mu^2\sum_\mu A_\mu^2\Big)
$$
For $c=-1$, the system is in the Coulomb phase. Exitations from the ground state
show that the vector particle (i.e.,\ the photon) has zero mass.

%-------------------------------------------------------------------------------
\section{The $U(2)$ Higgs Model.}

\newcommand{\fs}{\mbox{\footnotesize$\sqrt{2/3}$}}
\newcommand{\fr}{\mbox{\footnotesize$\sqrt{3}$}}
We now come to the $n=2$ situation. 
The $U(2)$ connection 1-form $A$ may be written in the basis $e_k$, $k=1,\ldots,4$ 
as introduced in Section 2:
$$
         A =\frac{i}{\sqrt{2}}\pmatrix{\sA+A^3 & A^1-iA^2\cr 
      A^1+iA^2&\sA -A^3\cr},\qquad A^k =dx^\mu A^k_\mu(x)
$$
where $A^k_\mu(x)$ ($k=1,\ldots,4$) are real gauge fields. A similar decomposition 
holds for the curvature $F=d_A^2$. Notice that $\tr A =i\fs A^4$.
We move on to write the superconnection $\DD$ (a $4\times 4$ matrix) in block form:
$$
            \DD =\pmatrix{ d+\hat{A}^+ &   i\Phi^* \cr i\Phi   & d+\hat{A}^-\cr },\qquad
           \hat{A}^+ =\pmatrix{0&0\cr 0 & i\fs A^4\cr}\ ,\qquad
           \hat{A}^- =A\ .
$$
Subgroups of $U(2)$ have a specific interpretation in the context of the electro-weak theory.
For instance, the  $U(1)$ subgroup consisting of phase transformations leads to the
conservation of the weak hypercharge in the unbroken phase.
The Higgs field $\Phi$ is some $2\times 2$-matrix whose columns represent Higgs doublets
in the fundamental representation of $SU(2)$: 
$$
                \Phi =\pmatrix{ \Phi_1 &\Phi_3\cr \Phi_2 &\Phi_4\cr}\ .
$$
The two doublets have opposite weak hypercharge. For an account of the physical implications
of two-doublet models see the review article by M.Sher [23].

There are five real second-order
forms , invariant under $U(2)$, that one may construct from the Higgs field:
\begin{eqnarray*}
   R_1(\Phi) &=&|\Phi_1|^2+|\Phi_2|^2,\qquad
   \phantom{R_4(\Phi)+i}R_3(\Phi)\ =\ |\Phi_1^*\Phi_3+\Phi_2^*\Phi_4| \\
   R_2(\Phi) &=&|\Phi_3|^2+|\Phi_4|^2,\qquad  
   R_4(\Phi)+iR_5(\Phi)\ =\ \Phi_1\Phi_4-\Phi_2\Phi_3
\end{eqnarray*}
Since they satisfy the relation
\begin{equation}
             R_1R_2 =R_3^2+R_4^2+R_5^2 \ , \label{man}
\end{equation}
only four of them are algebraically independent.
It is natural to think of the manifold (\ref{man}) as some  moduli space 
related to superconnections. 

In principle, any gauge invariant Higgs potential $V(\Phi)$, be it the classical or the
effective potential,  can be written as a function of the above invariants. Such a representation 
is convenient because the problem of minimizing the action is then reduced to solving a 
simpler problem in lower dimension. Each solution provides certain constants 
$$    r_{i}=R_i(\Phi_c),\quad i=1,\ldots 5, $$
which characterize the gauge orbit of $\Phi_c$, and the moduli space of vacua becomes a 
submanifold of
\begin{equation}
                     r_1r_2=r_3^2+r_4^2+r_5^2\ .
\end{equation}
Given the numbers $r_i$, the next step would be to determine the eigenvalues of the mass
matrix $m^2$ as defined by (\ref{Qa}). Since the matrix elements depend on the choice of
$\Phi_c$, hence on the point of the gauge orbit chosen, we better look at the characteristic 
polynomial of that matrix which is gauge invariant and thus can be written entirely in terms 
of the variables $r_i$. By a tedious but straightforward calculation one finds:
\begin{equation}
    {\rm det}(m^2-\lambda)=
   (r-\lambda)^2\big(\lambda^2-\qu r\lambda+\qu r_3^2 \big),\qquad
   (0\le 2r_3\le r)
\end{equation}
where $r=r_1+r_2=-\hf{\rm Tr\,}L_c^2$.
An immediate consequence is the following alternative:
\begin{quote}\em
    (1) If $r=0$, the eigenvalue zero of the mass matrix is fourfold degenerate: all
    vector bosons are massless.\\
    (2) If $r>0$, the eigenvalue $r$ of the mass matrix is twofold degenerate. We  
    interpret $r$ as the mass (squared) of the $W^{\pm}$ bosons.
    If $r_3=0$, there is an eigenvalue zero naturally associated to the photon and an
    eigenvalue $\qu r$ interpreted as the mass (squared) of the $Z$ boson
   so that $m_W^2:m_Z^2=3:4$.
   In the latter case, the residual gauge group is isomorphic to $U(1)$, leading
   to the notion of the electric charge. The $W^\pm$ bosons receive the charge
   $\pm 1$, while the $Z$ boson is neutral.
\end{quote}
So far we have not fixed the value of the mass parameter $\mu$. From now on we
shall always assume that $\mu^2 =r$ so that $\mu$ coincides with the W mass.
In other words, it is the W mass that sets the mass scale.
The $U(2)$ Higgs model conforms to the existing emperical data only
if $r_3=0$.  It is therefore important to show that $r_3=0$ is not an extra assumption 
but follows from the Higgs potential (\ref{HiPo}).
Though the Higgs potential depends on arbitrary constants $c_0$, $c_1$, and $c_2$, it is 
special among  $U(2)$ invariant fourth-order polynomials. For a simple calculation reveals that
\begin{equation}
         V(\Phi)=\big(R_1(\Phi)-b_1\big)^2+
                         \big(R_2(\Phi)-b_2\big)^2+2R_3(\Phi)^2+ b_3^2\ .
\end{equation}
Here we passed from the set $c_i$ of constants to another set $b_i$ given by
$$
         b_1=\hf\mu^2(c_0+c_1),\quad 
         b_2=\hf\mu^2(c_2+c_1),\quad
         b_3=\hf\mu^2\big( (c_0-c_1)^2+(c_2-c_1)^2\big)^{1/2}\ .
$$
While $b_3$ is physically irrelevant, $b_1$ and $b_2$ are essential for spontaneous
symmetry breaking. It is now obvious that any ground state has coordinates 
$$
          r_i=\max(0,b_i),\ (i=1,2),\quad r_3=0, \quad r_4^2+r_5^2=r_1r_2=\mbox{const.}
$$
which establishes two things: (1) our claim that $r_3=0$ and (2) the moduli space of vacua
is the sphere $S^1$ provided $r_1r_2>0$, or simply a point if either $r_1=0$ or $r_2=0$.
Granted the condition $r_3=0$ we can always perform a $U(2)$ gauge transformation so that the Higgs condensate
assumes the form
\begin{equation}
              \Phi_c=\pmatrix{r_1^{1/2}e^{i\alpha}  & 0\cr 0 & r_2^{1/2}\cr}\ ,\qquad
              r_1+r_2=\mu^2 ,\qquad
              r_4+ir_5=(r_1r_2)^{1/2}e^{i\alpha}.  \label{Phc}
\end{equation}
with $\alpha$ parametrizing the sphere $S^1$. The choice of such a standard form is
essential for getting a standard set of eigenvectors of the mass matrix.
In fact it follows at once from (\ref{Qa}) and (\ref{Phc}) that
$$
      Q(a) =\mu^2\big(|a^1+ia^2|^2+(\af+a^3)^2\big)\ ,
$$
and thus the eigenvectors of the mass matrix are:
$$\begin{tabular}{|c|c|}
    $\quad {\rm mass}^2\quad$ & eigenvector\\ \hline
    $\vphantom{\Big(}0$ & $\hf(\fr a^4-a^3)$\\[2mm]
    ${\textstyle\frac{4}{3}}\mu^2$&$\hf(\fr a^3+a^4)$\\[2mm]
    $\mu^2 $& $a^1,\, a^2$
\end{tabular}
$$
The residual gauge group, isomorphic to $U(1)$, is 
\begin{equation}
   G_0=\Big\{\pmatrix{1&0\cr 0&e^{i\alpha}\cr}\in U(2)\ \Big|\ 0\le\alpha<2\pi\Big\}
       \label{res}
\end{equation}
We relate the photon, the Z boson, and the W boson to the above eigenvectors and
thus work with the following fields:
\begin{equation}
              A^0 =\hf(\fr A^4- A^3),\qquad Z=\hf(\fr A^3+A^4),\qquad
              W^\pm={\textstyle\frac{1}{\sqrt{2}}}(A^1\mp iA^2)     \label{pho}
\end{equation}
It is common practice to write $Z=\cos\theta\, A^3+\sin\theta\, A^4$ with $\theta$ the
Weinberg angle and to determine $\sin^2\theta$ by experiment. Comparison with (\ref{pho})
shows that $\sin^2\theta=1/4$ in the present theory while the generally accepted value 
obtained from experiment is $\sin^2\theta=0.231$. The value $1/4$, however, has 
previously been predicted on different grounds (see for instance Refs.\ [6],
[8], and [18]). 

As a matrix, the connection 1-form may now be written in terms of $A^0$, $Z$ and $W$:
\begin{equation}
         A=i\pmatrix{\rf{2}{3}Z & W^+\cr W^-& \rf{1}{2}A^0-\rf{1}{6}Z\cr}
\end{equation}

To summarize, the choice of the central charges $c_i$ does not seem to matter as long as we 
keep $r$ at a fixed value, say $\mu^2$. That this impression is false will become clear as
soon as the coupling to matter is taken into account. In a forthcoming paper, the Yukawa
interaction of fundamental fermions with the Higgs field results from a widening of the
concept of Dirac operators  and so is viewed as integral part of the gauge coupling. It will
then become clear that the invariant parameters $r_1$ and $r_2$ are proportional to
the masses (squared) of the pair $(\nu_e,e)$ in a purely leptonic model 
(one generation only). Vanishing of the neutrino mass requires that $r_1=0$. 
In this situation, the moduli space of vacua shrinks to a point. This puts another constraint
on the parameters $c_i$, namely $c_0+c_1\le 0$ or $b_1\le 0$.

Let us now discuss the Higgs field itself.  The assignment of electric charges to
the four complex degrees of freedom can be read off from $\pmatrix{0&+1\cr -1&0}$.
The one-doublet model of Salam and Weinberg is recovered if 
\begin{equation}
                       \Phi=\pmatrix{0&\Phi^+\cr 0&\Phi^0\cr}
\end{equation}
which is consistent with $r_1=0$ but not with $r_1>0$.

The two-doublet model we propose has eight real degrees of freedom. Three of them 
can be gauged away, giving an extra polarisation degree of freedom to each massive
gauge field. The remaining five degrees can be arranged as follows:
$$
     \Phi=\pmatrix{X^0&0\cr  X^- &\rf{1}{2}\phi\cr}+\Phi_c
$$  
There are two complex fields $X^0$ and $X^-$. The real field $\phi$ describes the
Higgs particle of the conventional theory. To determine the (bare) masses of these fields we
expand the Higgs potential to second order assuming $r_1=0$:
$$
    V(\Phi)=b_1^2+b_3^2-2b_1|X^0|^2+2(\mu^2-b_1)|X^-|^2+
                    2\mu^2\phi^2+\ldots
$$
Recall now that $2b_1=\mu^2(c_0+c_1)<0$. Therefore,
$$
    m^2_{X^0} =\mu^2|c_0+c_1|,\quad m^2_{X^-}=\mu^2(2+|c_0+c_1|),\quad
    m^2_H =4\mu^2\ .
$$
In this scenario, the hypthetical X particles have masses that depend on the constants
$c_i$ while the mass of the Higgs particle is not influenced by their values. We may state
our results as
\begin{equation}
                   m_{\rm W}^2:m_{\rm Z}^2:m_{\rm H}^2=3:4:12,\quad
                   m^2_{X^-}=m^2_{X^0}+2m^2_W                  
\end{equation}
with the prediction $m_H=2m_W=$161 GeV. A value of the Higgs mass near 160 GeV
has also been predicted by Okumura [24].

To summarize, gauge potentials, Higgs fields, and the Higgs condensate can be accommodated
in a single Hermitian $4\times 4$ matrix:
\begin{equation}
     \hat{A}+L=i\pmatrix{
      0             &                0                              &   \bar{X}^0      &    \bar{X}^-                \cr
      0             &\rf{1}{2}A^0+\rf{1}{6}Z  &       0                   & \rf{1}{2}\phi +\mu    \cr                         
      X^0       &       0                                      &\rf{2}{3}Z           & W^+                             \cr 
      X^-       &  \rf{1}{2}\phi +\mu            &  W^-                 &\rf{1}{2}A^0-\rf{1}{6}Z \cr}
\end{equation}
The electric charges $0,\pm 1$ attributed to the entries of such a matrix may be read off from
the scheme:
$$
\pmatrix{0 &+1  & 0 & +1\cr  -1 &0 & -1 & 0\cr 
                0 &+1  & 0 & +1\cr  -1 &0 & -1 & 0\cr}   $$
\par\noindent
It should be kept in mind that we rely here on a classical approximation. Quantization
changes the Higgs potential to some effective potential which is expected to
considerably differ from the classical potential. The same proviso applies to the computation 
of masses since they also depend on the effective potential. 
To include loop corrections is one way
to change predictions, perhaps not in a reliable way. Such corrections depend
on the mass matrices of matter fields and thus are outside the scope of this paper.
Another way is to apply renormalization group methods which also rely on loop calculations.
It should also be kept in mind that the relation $\tan\theta=g'/g$ holds for $g$ and $g'$
defined on a sliding energy scale. Therefore, the Weinberg angle $\theta$ cannot be a constant 
over a large energy range. The values to be used here should come from energies comparable to the mass parameter 
$\mu$.

The superconnection formalism can be extended to include matter fields as will be shown
in a subsequent paper.
It would also be desirable to push the theory further, so as to obtain a unified theory of weak,
electromagnetic, and strong interactions as a gauge theory based on a larger group
incorporating both the vector bosons of the electroweak theory and the 
gluons of QCD. It is not clear at the moment whether such an approach will give
reasonable results.

{\bf Acknowledgements.} The author wishes to thank Stephan Korden and Thomas Strobl
for numerous discussions.

\newpage

\leftline{\Large \bf References}\vspace{2mm}
\begin{enumerate}
\item S.\ Weinberg, Rev.\ Mod.\ Phys.\ {\bf 92}, 515 (1980)\newline
          S.\ Weinberg: The Quantum Theory of Fields, Vol.II: Modern Applications.
         Cambridge University Press 1996
\item K.\ B.\ Marathe, G.\ Martucci: The Mathematical Foundation of Gauge Theories,
          North-Holland 1992
\item C.\ Nash: Differential Topology and Quantum Field Theory, Academic Press 1991
\item E.\ Witten, Phys.\ Rev.\ Lett.\ {\bf 38}, 121 (1977)
\item N.\ Manton, Nucl.\ Phys.\ {\bf  B 158}, 141 (1979)
\item D.B.\ Fairlie, Phys.\ Lett.\ {\bf B 82}, 97 (1979)
\item P.\ Forg\'acs, N.\ Manton, Comm.\ Math.\ Phys. {\bf 72}, 15 (1980)\newline
         J.\ Harnad, S.\ Shnider, J.\ Tafel, Lett.\ Math.\ Phys.\ {\bf 4}, 107 (1980)
\item Y.\ Ne'eman, Phys.\ Lett.\ {\bf B 81}, 190 (1979)
\item Y.\ Ne'eman, S.\ Sternberg, Proc.\ Nat.\ Acad.\ Sci.\ USA {\bf 87}, 7875 (1990)
\item Y.\ Ne'eman, hep-th/9708052
\item A.\ Connes, Non-commutative Geometry, Academic Press, New York 1994
\item F.\ Scheck, The Standard Model within Non-commutative Geometry: A
         Comparison of Models, hep-th/9701073, and references therein.
\item A.\ Connes, J.\ Lott, Nucl.\ Phys.\ {\bf B18}(Proc.\ Suppl.), 29 (1990);
          Proc.\ Carg\`ese Summer School 1991 (eds.\ J.\ Fr\"ohlich et al.), Plenum Press
          New York 1992
\item Y.\ Okumura, Prog.\ Theor.\ Phys.\ {\bf 65}, 969 (1996)
\item S.\ Balakrishna, F.\ G\"ursey, K.\ C.\ Wali, Phys.\ Lett.\ {\bf B254}, 430 (1991)
   \newline
           R.\ Erdem,  hep-th/9701067
\item D.\ Quillen, Topology {\bf 24}, 89 (1985)
\item V.\ Mathai, D.\ Quillen, Topology {\bf 25}, 85 (1986); see also:\newline
          N.\ Berline, E.\ Getzler, M.\ Vergne: Heat Kernels and Dirac Operators,
          Springer, Berlin 1996
\item R.\ Coquereaux, Higgs Fields and Superconnections, Lecture Notes in Physics, Vol.\ 375
          \newline
		    R.\ Coquereaux,  G.\ Esposito-Farese, G.\ Vailant, Nucl.\ Phys.\ {\bf B353}, 689
           (1991)
\item C.-Y.\ Lee. hep-th/9602159\newline
\item H.\ Figueroa, J.\ M.\ Gracia-Bond\'{i}a, F.\ Lizzi, J.\ C.\ V\'{a}rilly, hep-th/9701179
\item C.\ H.\ Taubes, Commun.\ Math.\ Phys.\ {\bf 75}, 207 (1980)
\item M.\ Blau, J.\ Geom.\ Phys.\ {\bf 11}, 95 (1993)
\item M.\ Sher, Phys.\ Reports {\bf 179}, 273 (1989)
\item Y.\ Okumura, hep-ph/9707350
\end{enumerate}
\end{document}